# On the relative role of different age groups during influenza A epidemics in Germany, 2002-2017

Edward Goldstein [1]

1. London, UK. Email: edmugo3@gmail.com

## Abstract

Background: There is limited information about the role of different age groups, particularly subgroups of school-age children and younger adults in propagating influenza epidemics.

Methods: For a communicable disease outbreak, some subpopulations may play a disproportionate role during the ascent of the outbreak due to increased susceptibility and/or contact rates. Such subpopulations can be identified by considering the proportion that cases in a subpopulation represent among all cases in the population occurring before the epidemic peak (Bp), the corresponding proportion after the epidemic peak (Ap), to calculate the relative risk for a subpopulation, RR=Bp/Ap. We estimated RR for several age groups using data on reported influenza A cases in Germany between 2002-2017.

Results: Children aged 14-17y had the highest RR estimates for 7 out of 15 influenza A epidemics in the data, including the 2009 pandemic, and the large 2016/17, 2008/09, and 2006/07 seasons. Children aged 10-13y had the highest RR estimates during 3 epidemics, including the large 2014/15 and 2004/05 seasons. Children aged 6-9y had the highest RR estimates during two epidemics, including the large 2012/13 season. Children aged 2-5y had the highest RR estimate during the moderate 2015/16 season; adults aged 18-24y had the highest RR estimate during the small 2005/06 season; adults aged 25-34y had the highest RR estimate during the large, 2002/03 season.

Conclusions: Our results support the prominent role of all school-age children, particularly the oldest ones, in propagating influenza epidemics in the community. We note that national vaccination coverage levels among older school-age children were lower than among younger school-age children during the recent influenza seasons in the US, and influenza vaccination program in England has not been phased in yet for secondary school students.

**Introduction**

Despite various advances in our understanding of influenza epidemiology, there is still uncertainty about the role of different age groups in propagating influenza epidemics in the community. That uncertainty is related to the lack of consensus on how to quantify the roles of different population groups during an epidemic, to the granularity of our understanding of those roles in terms of age splitting (e.g. the relative prominence of different subgroups of school-age children), and to the consistency of those roles during different influenza seasons in different locations. Serological studies have suggested that school-age children experienced the highest influenza infection rates during the 2009 A/H1N1 pandemic (e.g. [1,2]) and during certain seasonal influenza epidemics [3], though for other influenza seasons, infection rates were found to be similar across different age groups [4]. Studies based on transmission-modeling (e.g. [5-7]) point to the prominence of school-age children in propagating influenza epidemics. Those studies represent an important advance towards our understanding of influenza transmission in the community, and the evaluation of the impact of influenza vaccination. At the same time, due to the complexity of influenza dynamics in the community there is uncertainty in the assumptions made in those transmission models; moreover, issues such as seasonal variability in the relative role of different subgroups of school-age children and young adults during influenza outbreaks have received limited coverage in transmission-modeling studies calibrated against epidemic data, particularly in the non-pandemic context.

In [8], we introduced a method for evaluating the role of different age groups during influenza outbreaks. The idea of that method is that subpopulations that play a disproportionate role during the epidemic's ascent (due to increased susceptibility and/or contact rates) can be identified via the relative risk (RR) statistic that assesses the change in the subpopulation's proportion among cases in the whole population before vs. after the epidemic's peak [8,9]. In [8], we have estimated the RR statistic for several epidemics associated with different influenza (sub)types in the US between 2009-2014 using data on hospitalized cases. An important aspect of the method proposed in [8] is that the results do not depend on case-hospitalization rates in different age groups. Moreover, we have used simulations to show a relation between a higher value for an RR statistic in a given age group and a higher impact of vaccinating an individual in that age group on reducing the epidemic's initial growth rate.

While the results in [8] suggested the prominence of school-age children in propagating the major influenza A epidemics, our data were not sufficiently granular to evaluate the role of different age subgroups of school-age children, and of younger adults during those epidemics. In this paper, we apply the RR statistic to data from 15 influenza A epidemics in Germany (the 2002/03 through the 2016/17 seasons). Availability of weekly data on influenza A cases stratified by year of age from the Robert Koch Institute [10] allowed for an examination of the role of different age groups for a finer age splitting than in [8] across a larger number of epidemics.

**Methods**

*Data*
We utilized weekly data on influenza A cases between 2002-2017 stratified by year of age. We included all cases that were reported according to the Communicable Diseases Law (IfSG), with clinical and laboratory criteria being met [10]. While there is some variability in the age splitting for the school system in different parts of Germany, we have chosen age groups for children that generally reflect the kindergarten, primary, and secondary school age groups, with the latter further split into two. The ten age groups that we have considered for our analyses are (0-1y,2-5y,6-9y,10-13y,14-17y,18-24y,25-34y,35-49y,50-64y,65+y).

*Statistical analyses*
Each influenza season was defined as a period starting from calendar week 40 of a given year and ending on calendar week 20 of the following year. Due to atypical influenza circulation patterns during the 2009 A/H1N1 pandemic, the 2008/09 season was defined to end on calendar week 16 of 2009, and the 2009/10 season was defined to start on week 37, 2009, and end on week 10, 2010 (we have not included cases reported during the summer of 2009 in our analyses due to changes in contact patterns following school opening in 2009). The 2016/17 season was defined to end on week 16, 2017.

The weekly counts for the number of reported cases that meet the criteria described in the Data section for the 15 seasons used in our study are plotted in Figure 1. We note that those data do not necessarily reflect the relative magnitudes of influenza attack rates during different seasons due to a variety of issues such as the apparent upward trend in reporting, potential changes in reporting patterns associated with the 2009 A/H1N1 epidemic, etc. One can get a crude assessment of the magnitudes of different influenza A seasons in Germany based on surveillance data, particularly the examination of Figure 20 in [11], Figure 10 in [12], Table 1 in [12], and Figures 1 and 3 in [13]. That comparison suggests that the 2002/03, 2004/05, 2006/07, 2008/09,

2009/10, 2012/13, 2014/15, and 2016/17 seasons had the larger influenza A epidemics, with the 2015/16 season having a more moderate influenza A epidemic.

For each season $s$, we defined the peak week $P(s)$ as the calendar week with the highest total number of reported influenza A cases. Each case during the given season was classified as before-the-peak case if it occurred in or prior to week $P(s)-2$; after-the-peak case if it occurred in or after week $P(s)+2$. We excluded cases occurring between weeks $P(s)-1$ and $P(s)+1$ from the analysis to avoid misclassification of counts as before or after-the-peak stemming from the noise in the count data for the reported cases, differences in *case-reporting rates* (proportion of influenza A infections in the population that are reported according to the criteria described in the Data section) for the various age groups that may result in different peak times for the counts of the reported cases vs. cases of infection in the community, etc.

For a given season, for each age group $g$, let $B(g)$ and $A(g)$ denote the before-and-after the peak counts of reported cases in that age group. The point estimate for the seasonal relative risk $RR(g)$ in an age group $g$ is

$$\frac{B(g)}{\sum_h B(h)} \bigg/ \frac{A(g)}{\sum_h A(h)} \quad (1)$$

The estimates and confidence bounds for relative risks in each group were obtained in a Bayesian framework following the methodology in [8]. Briefly, we assume that the numbers of cases of infection in the population in each age group are large and the case-reporting rates are low, thus the numbers of reported cases in each age group before and after the peak are well-approximated by Poisson variables (if case-reporting rates in certain age groups during certain seasons are not too low, the Poisson approximation overestimates the variance in the numbers of reported cases). Samples $B^i(h), A^i(h)$ (i=1,..,100000) for the Poisson parameters corresponding to the counts $B(h), A(h)$ (with a flat prior) are generated; for each i=1,..,100000, the corresponding parameters $B^i(h), A^i(h)$ are plugged into eq. 1 to generate an estimate $RR^i(g)$ for the relative risk. The mean and the credible interval for the sample $\left(RR^i(g)\right)$ (i=1,..,100000) are then extracted.

**Results**

Figure 1 plots the weekly counts for the number of reported cases that meet the criteria described in the Data subsection of the Methods for the 15 seasons used in our study.

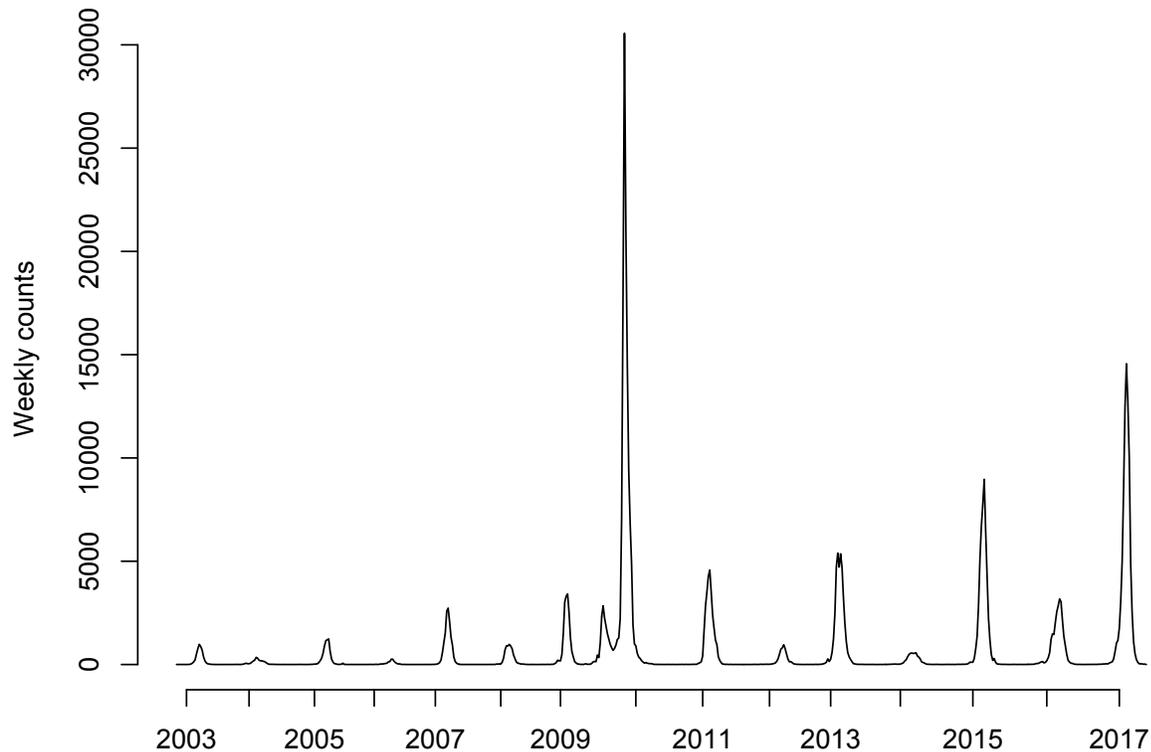

**Figure 1:** Weekly counts for the number of reported influenza A cases for the 2002/03 through the 2016/17 seasons.

Tables 1 and 2 exhibit the RR estimates for children and adults correspondingly for the 2002/03 through the 2016/17 influenza A epidemics in Germany. Estimates of the RR statistic were generally the highest for school-age children, particularly the older ones. Children aged 14-17y had the highest RR estimates for 7 out of 15 influenza A epidemics in the data, including the 2009 pandemic, and the large 2016/17, 2008/09, and 2006/07 seasons. Children aged 10-13y had the highest RR estimates during 3 epidemics, including the large 2014/15 and 2004/05 seasons. Children aged 6-9y had the highest RR estimates during two epidemics, including the large 2012/13 season. Children aged 2-5y had the highest RR estimate during the moderate 2015/16 influenza A season; adults aged 18-24y had the highest RR estimate during the small 2005/06 season; adults aged 25-34y had the highest RR estimate during the large, 2002/03 season.

| Season / Age group | 0-1y | 2-5y | 6-9y | 10-13y | 14-17y |
|---|---|---|---|---|---|
| 2002/03 | 0.82(0.62,1.05) | 1.13(0.98,1.29) | 1.22(0.94,1.55) | 0.96(0.73,1.23) | 1.3(0.97,1.72) |
| 2003/04 | 1.25(0.96,1.59) | 0.81(0.66,0.98) | 0.88(0.62,1.18) | 1.23(0.94,1.57) | 1.61(1.2,2.1) |
| 2004/05 | 0.58(0.44,0.76) | 0.9(0.75,1.07) | 1.71(1.36,2.14) | 1.87(1.35,2.57) | 1.37(0.95,1.94) |
| 2005/06 | 0.73(0.4,1.2) | 0.85(0.65,1.1) | 1.15(0.8,1.59) | 1.07(0.66,1.66) | 1.12(0.6,1.94) |
| 2006/07 | 0.69(0.57,0.81) | 1.04(0.97,1.12) | 1.16(1.06,1.28) | 0.99(0.87,1.13) | 1.33(1.08,1.63) |
| 2007/08 | 0.56(0.43,0.72) | 1.12(1.01,1.25) | 1.27(1.14,1.41) | 1.37(1.15,1.63) | 0.98(0.75,1.24) |
| 2008/09 | 0.87(0.76,1) | 1.24(1.15,1.35) | 1.24(1.08,1.41) | 1.36(1.15,1.59) | 1.98(1.54,2.53) |
| 2009/10 | 0.35(0.3,0.41) | 0.45(0.42,0.49) | 0.71(0.67,0.74) | 1.38(1.32,1.43) | 1.83(1.76,1.91) |
| 2010/11 | 0.85(0.76,0.96) | 1.07(1,1.15) | 1.47(1.34,1.6) | 1.34(1.2,1.49) | 1.39(1.21,1.59) |
| 2011/12 | 0.85(0.67,1.08) | 1.08(0.96,1.22) | 1.26(1,1.58) | 1.25(0.89,1.73) | 1.3(0.86,1.9) |
| 2012/13 | 1.26(1.13,1.39) | 1.57(1.47,1.67) | 1.83(1.67,2.01) | 1.49(1.31,1.68) | 1.35(1.17,1.55) |
| 2013/14 | 1.03(0.79,1.34) | 1.35(1.14,1.58) | 1.12(0.87,1.42) | 0.99(0.7,1.4) | 1.5(1.03,2.18) |
| 2014/15 | 0.86(0.77,0.95) | 1.19(1.12,1.27) | 1.59(1.44,1.74) | 1.84(1.64,2.05) | 1.45(1.29,1.62) |
| 2015/16 | 1.14(1.01,1.28) | 1.51(1.39,1.64) | 1.47(1.27,1.69) | 1.21(0.99,1.47) | 1.07(0.87,1.31) |
| 2016/17 | 0.93(0.84,1.03) | 1.01(0.94,1.08) | 1.64(1.5,1.78) | 2.06(1.88,2.25) | 2.28(2.08,2.5) |

**Table 1:** Seasonal RR estimates for the different age subgroups of children

| Season / Age group | 18-24y | 25-34y | 35-49y | 50-64y | 65+y |
|---|---|---|---|---|---|
| 2002/03 | 1.05(0.69,1.53) | 1.5(1.12,1.96) | 1.11(0.89,1.38) | 0.78(0.57,1.04) | 0.19(0.11,0.3) |
| 2003/04 | 0.99(0.64,1.44) | 0.92(0.6,1.33) | 0.78(0.55,1.07) | 0.9(0.57,1.33) | 0.72(0.38,1.21) |
| 2004/05 | 1.43(0.88,2.29) | 1.03(0.75,1.4) | 0.92(0.73,1.15) | 0.81(0.6,1.08) | 0.33(0.25,0.44) |
| 2005/06 | 2.35(1.15,4.47) | 0.96(0.57,1.5) | 1.07(0.76,1.47) | 1.03(0.6,1.64) | 0.97(0.52,1.66) |
| 2006/07 | 0.94(0.7,1.25) | 0.96(0.76,1.19) | 0.9(0.76,1.06) | 0.8(0.62,1.01) | 0.67(0.49,0.88) |
| 2007/08 | 0.73(0.51,1.01) | 0.62(0.46,0.82) | 0.79(0.66,0.93) | 0.54(0.38,0.75) | 0.43(0.21,0.74) |
| 2008/09 | 0.91(0.75,1.1) | 0.82(0.71,0.95) | 0.77(0.68,0.86) | 0.58(0.49,0.67) | 0.64(0.5,0.8) |
| 2009/10 | 1.21(1.15,1.27) | 0.93(0.88,0.99) | 0.88(0.84,0.93) | 0.67(0.62,0.73) | 0.44(0.34,0.54) |
| 2010/11 | 1.04(0.94,1.15) | 0.92(0.85,1) | 0.89(0.83,0.95) | 0.68(0.62,0.74) | 0.44(0.35,0.54) |
| 2011/12 | 0.81(0.53,1.19) | 1.13(0.85,1.47) | 1.08(0.9,1.29) | 0.97(0.78,1.19) | 0.67(0.55,0.8) |
| 2012/13 | 0.88(0.76,1.01) | 0.95(0.87,1.04) | 0.89(0.83,0.95) | 0.61(0.56,0.66) | 0.28(0.23,0.32) |
| 2013/14 | 0.95(0.67,1.31) | 0.97(0.78,1.21) | 1.11(0.96,1.3) | 0.77(0.66,0.91) | 0.65(0.53,0.79) |
| 2014/15 | 1.06(0.93,1.2) | 0.96(0.89,1.04) | 0.95(0.9,1.01) | 0.81(0.76,0.86) | 0.67(0.63,0.71) |
| 2015/16 | 1.05(0.89,1.24) | 1.05(0.96,1.15) | 1.03(0.97,1.1) | 0.79(0.74,0.84) | 0.53(0.48,0.58) |
| 2016/17 | 1.35(1.24,1.47) | 0.88(0.82,0.94) | 1.02(0.97,1.07) | 0.84(0.81,0.87) | 0.77(0.74,0.79) |

**Table 2:** Seasonal RR estimates for the different age subgroups of adults

**Discussion**

Our understanding about the roles of different age groups during influenza epidemics, and the seasonal variability in those roles is still limited. In this paper we aim at contributing towards that understanding by examining the relative role of individuals in different age groups during 15 influenza A epidemics in Germany using the previously developed methodology [8,9]. Our results suggest the prominence of school-age children, particularly the older ones, during the larger influenza A epidemics. The relative contributions of children aged 2-5y and younger adults were generally more modest, and the relative contributions of children aged 0-1y and individuals over the age of 50y were generally even smaller.

Our study has some limitations. The relation between the RR statistic that we have utilized and the role played by an average individual in a given age group during an influenza outbreak is not entirely clear. Our earlier work ([8]) had attempted to address this issue through simulations of transmission dynamics, finding an association between the RR statistic and the impact of vaccination on the epidemic's initial growth rate/reproductive number. Usage of national peaks of reported influenza A cases in Germany may result in mischaracterization of cases as occurring before-vs.-after the peak for the corresponding local epidemics due to potential spatial asynchrony in influenza epidemics within Germany. We note that this would bias all the RR estimates toward the null value of 1, which shouldn't take away from the conclusions about the prominence of certain population groups during different epidemics. Figure 1 suggests temporal changes in reporting practices; however this should not bias the RR estimates that are derived separately during each season, unless reporting practices change through the course of the season in a manner that is not uniform for all age groups [8,9].

We believe that despite those limitations our study sheds new light on the relative contribution of individuals in different age groups, particularly subgroups of children and younger adults towards propagating influenza epidemics. Our results point to the prominence of school-age children, particularly the older ones, during the larger influenza A outbreaks. During the recent influenza seasons, national vaccination coverage levels among older school-age children were lower than among younger school-age children in the US [14], and influenza vaccination program in England has not been phased in yet for secondary school students [15]. We hope that our findings would contribute to the debate about the utility of vaccination of all school-age children towards mitigating influenza epidemics in the whole community.

**Acknowledgement:** The data used in our analyses were made available by the Robert Koch Institute.